\documentclass[10pt]{article}

\usepackage{amsmath}
\usepackage{amssymb}

\usepackage{graphicx,multirow}
\usepackage[hang,tight]{subfigure}

\usepackage{cite,url}

\usepackage{color}

\usepackage[labelfont=bf,labelsep=period,justification=raggedright]{caption}

\makeatletter
\renewcommand{\@biblabel}[1]{\quad#1.}
\makeatother

\date{}

\begin{document}
\vspace*{0.35in}

\begin{flushleft}
{\Large
\textbf{Club Convergence of House Prices: Evidence from China's Ten Key Cities}
}
\newline
\\
\bigskip
Hao Meng\textsuperscript{1,2},
Wen-Jie Xie\textsuperscript{2,3,4},
Wei-Xing Zhou\textsuperscript{1,2,4,*},

\bigskip

\bf{1} Department of Mathematics, School of Science, East China University of Science and Technology, Shanghai 200237, China
\\
\bf{2} Research Center for Econophysics, East China University of Science and Technology, Shanghai 200237, China
\\
\bf{3} Postdoctoral Research Station, East China University of Science and Technology, Shanghai 200237, China
\\
\bf{4} Department of Finance, School of Business, East China University of Science and Technology, Shanghai 200237, China
\\
\bigskip

* E-mail: wxzhou@ecust.edu.cn (WXZ)

\end{flushleft}

\section*{Abstract}

The latest global financial tsunami and its follow-up global economic recession has uncovered the crucial impact of housing markets on financial and economic systems. The Chinese stock market experienced a markedly fall during the global financial tsunami and China's economy has also slowed down by about 2\%-3\% when measured in GDP. Nevertheless, the housing markets in diverse Chinese cities seemed to continue the almost nonstop mania for more than ten years. However, the structure and dynamics of the Chinese housing market are less studied. Here we perform an extensive study of the Chinese housing market by analyzing ten representative key cities based on both linear and nonlinear econophysical and econometric methods. We identify a common collective driving force which accounts for 96.5\% of the house price growth, indicating very high systemic risk in the Chinese housing market. The ten key cities can be categorized into clubs and the house prices of the cities in the same club exhibit an evident convergence. These findings from different methods are basically consistent with each other. The identified city clubs are also consistent with the conventional classification of city tiers. The house prices of the first-tier cities grow the fastest, and those of the third- and fourth-tier cities rise the slowest, which illustrates the possible presence of a ripple effect in the diffusion of house prices in different cities.

\section*{Introduction}

The U.S. housing market experienced a continuous rise since the 1990's, which was driven by diverse factors such as the wealth effect and the inflow of international capitals \cite{Sornette-Zhou-2004-PA}. According to the log-periodic power-law model \cite{Sornette-2003,Sornette-2003-PR}, no bubble was detected in the US housing market in 2003 \cite{Zhou-Sornette-2003a-PA}. However, in 2005, evident signatures of a housing bubble were identified \cite{Zhou-Sornette-2006b-PA}, and a strikingly accurate forecast was released in the Abstract of Ref.~\cite{Zhou-Sornette-2006b-PA} stating that: ``From the analysis of the S\&P 500 Home Index, we conclude that the turning point of the bubble will probably occur around mid-2006.'' The turndown of the US house prices measured by the S\&P Case-Shiller house price index was indeed fulfilled in 2006, which triggered the outbreak of the US subprime mortgage crisis in 2007. The aftermath was very severe. It caused the credit crisis in the US and a national crisis in US's financial markets. The US financial crisis diffused to the worldwide financial markets and hastened a global financial crisis in 2008. In this case, stock markets acted perfectly as the barometer of real economies and a global economic recession followed unavoidably. What followed further was the European sovereign debt crisis. The worldwide economies are still struggling on the way to recover. The outline of this story demonstrates the crucial role played by an economy's housing market.

In the past three decades, China's economy experienced an unprecedented growth with an average growth rate of about 10\% and the capitalization of the Chinese stock market has become one of the largest all over the world. During the global financial tsunami, the Chinese stock market bubble bust and dropped by about 80\% with the Shanghai Stock Exchange Composite Index plummeted from its historical high at 6124 on 16 October 2007 to 1664 on 28 October 2008 \cite{Jiang-Zhou-Sornette-Woodard-Bastiaensen-Cauwels-2010-JEBO}, and China's economy has also slowed down by about 2\%-3\% when measured in GDP. Nevertheless, the housing markets in diverse Chinese cities seemed to continue the almost nonstop mania for more than ten years. In late 2003, the Shanghai House Price Composite Index has exhibited signatures of an undoubtable bubble in store \cite{Zhou-Sornette-2004a-PA}. In early 2008, the Shanghai housing market dropped mildly and then continued to soar, which was partially fuelled by the government bailout of 40 trillion Chinese yuan in November 2008.

There are still debates on whether there is a housing bubble in China. However, the consensus is apt to the presence of a bubble and many people think or hope that the bubble will crash sooner or later, although there are also many people deny the possibility of bubble burst, including some officials and house builders. Nevertheless, the possibility that the housing market will crash nationwide is a sword hanged on the development of China's economy. A crash of the housing market will cause severe damages to the economy and even cause social problems.

It has been well recognized that studying complex economic and financial systems under the framework of complex networks has crucial scientific significance, because the units or agents in complex systems interact with each other in a nonlinear manner \cite{Schweitzer-Fagiolo-Sornette-VegaRedondo-Vespignani-White-2009-Science}. In this work, we will investigate the correlation structure of house price indexes of 10 key cities in China based on both linear and nonlinear econophysical and econometric methods, which is closely related to the systemic risk of the national housing market \cite{Kritzman-Li-Page-Rigobon-2011-JPM,Billio-Getmansky-Lo-Pelizzon-2012-JFE,Zheng-Podobnik-Feng-Li-2012-SR,Meng-Xie-Jiang-Podobnik-Zhou-Stanley-2014-SR}. We identify city clubs and club convergence in the house price indexes and high systemic risk in the national housing market.

\section*{Materials and Methods}

\subsection*{Data sets}

\begin{figure}[htbp]
  \centering
  \includegraphics[width=0.9\linewidth]{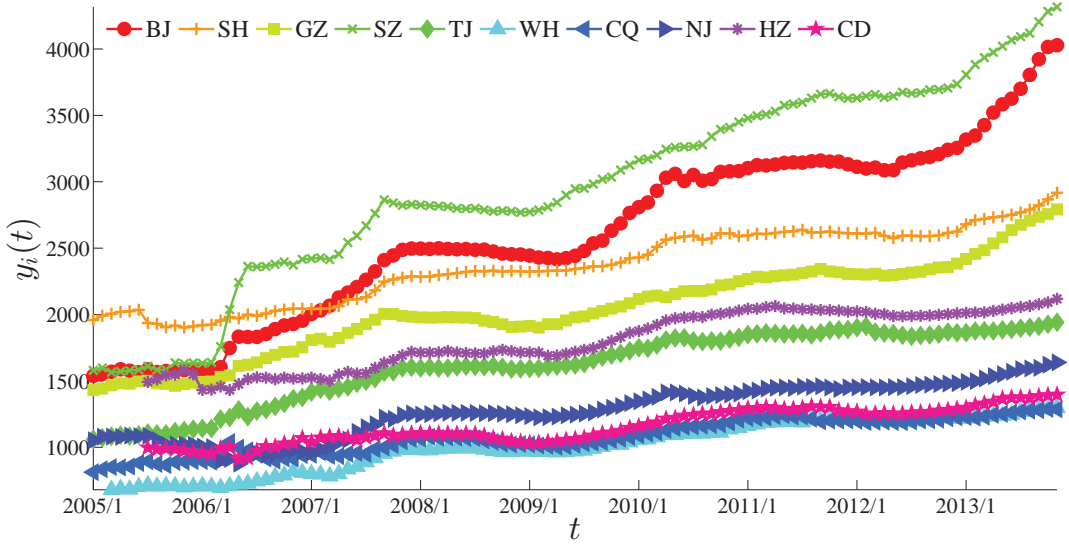}
  \caption{{\textbf{Evolution of the Urban Comprehensive Index of 10 key cities in China.}} All the indexes have risen during the time period under investigation.}
  \label{Fig:RawData}
\end{figure}

We use the monthly house price composite index (HPI) data for 10 key cities of China covering the period from January 2005 to November 2013, which were retrieved from the China Real Estate Index System (CREIS) of the China Index Academy. The data are publicly available at http://fdc.fang.com/index/XinFangIndex.aspx. The 10 key cities include Beijing (BJ), Shanghai (SH), Guangzhou (GZ), Shenzhen (SZ), Tianjin (TJ), Wuhan (WH), Chongqing (CQ), Nanjing (NJ), Hangzhou (HZ), Chengdu (CD). The house price indexes are constructed as the Urban Comprehensive Index which takes houses of residence, office edifice and commercial shop into account. The HPIs of these key cities are regarded as the vane of China's real estate market. Figure \ref{Fig:RawData} illustrates the HPI time series $y_i(t)$ ($t=1,2,\cdots,T$) of city $i$. Because the data of the first 6 months are unavailable for Wuhan, Hangzhou and Chengdu, we investigate the HPIs since July 2005, containing 101 data points for each city.

\subsection*{Correlation matrix and random matrix theory}

The random matrix theory (RMT) has been long applied in the econophysics community \cite{Laloux-Cizean-Bouchaud-Potters-1999-PRL,Plerou-Gopikrishnan-Rosenow-Amaral-Stanley-1999-PRL,Drozdz-Grummer-Gorski-Ruf-Speth-2000-PA,Plerou-Gopikrishnan-Rosenow-Amaral-Guhr-Stanley-2002-PRE,Tumminello-Lillo-Mantegna-2010-JEBO}. The similarity between two time series $y_i(t)$ and $y_j(t)$ is commonly calculated by the Pearson correlation coefficient as follows:

\begin{equation}
C_{ij} = \frac{\langle (y_i(t) - \langle y_i(t) \rangle)(y_j(t) - \langle y_j(t) \rangle) \rangle}{\sigma_i\sigma_j}.
\label{Eq: Correlation Coefficient}
\end{equation}
where $\langle \rangle$ is the calculation of mean and $\sigma_i$ is the standard deviation of time series $y_i(t)$. We study the raw correlation matrix $\mathbf{C}$, whose elements $C_{ij}$ are the Pearson correlation coefficients between various pairs of time series $y_i$ and $y_j$.

The eigenvalues and eigenvectors of $\mathbf{C}$ provide important information. In terms of the principal component analysis, for $N$ time series, the eigenvectors $v_i$ ($i=1,2,\ldots,N$) of the correlation matrix $\mathbf{C}$ are a full set of orthogonal axis in space which could decompose the total variability of all the time series into several orthogonal sub-variabilities by projecting observations on the axis. This set of decompositions suggests that the variability summarized by the first (largest) eigenvalue $\lambda_1$ and its corresponding eigenvector $v_1$ is the maximum among all possible orthogonal choice of the axis, the second largest eigenvalue $\lambda_2$ and its corresponding eigenvector $v_2$ then summarize the maximum variation in the unexplained portion of the original series after excluding the information explained by $\lambda_1$, and so on up to the smallest eigenvalue $\lambda_N$. The percent of variability explained by projecting observations on each eigenvector $v_i$ can be calculated as follows:
\begin{equation}
  \varphi_{i}=\frac{\lambda_i}{\sum_{k=1}^{N}\lambda_k}=\frac{\lambda_i}{N}.
  \label{Eq:ContributionRatio}
\end{equation}
We could also calculate the cumulative percent up to the $i$th eigenvalue as follow:
\begin{equation}
  \phi_{i}=\frac{\sum_{k=1}^{i}\lambda_k}{\sum_{k=1}^{N}\lambda_k},
  \label{Eq:CumulativeContributionRatio}
\end{equation}
which is also called the absorption ratio \cite{Kritzman-Li-Page-Rigobon-2011-JPM} and is a measure of systemic risk \cite{Billio-Getmansky-Lo-Pelizzon-2012-JFE}.

Applications of the random matrix theory (RMT) to stock market \cite{Plerou-Gopikrishnan-Rosenow-Amaral-Guhr-Stanley-2002-PRE,Song-Tumminello-Zhou-Mantegna-2011-PRE} and housing market \cite{Meng-Xie-Jiang-Podobnik-Zhou-Stanley-2014-SR} show that, the largest eigenvalue $\lambda_1$ and its corresponding eigenvector $v_1$ characterized the collective response of the entire market to a common stimuli. If there is a strong collective behavior in the market, all components would participate almost identically in the $v_1$, representing an influence that is common to all stocks. $\lambda_1$ and its corresponding $\varphi_1$ would be extremely large, interpreting most of the variability in the observations. One could further unravel some grouping information from other largest eigenvalues and their associated eigenvectors which deviate from the RMT predictions. Besides, the smallest eigenvalue $\lambda_N$ and its corresponding eigenvector $v_N$ could highlight pairs of stocks with a correlation coefficient much larger than the average, namely ``decoupling'' from other stocks \cite{Plerou-Gopikrishnan-Rosenow-Amaral-Guhr-Stanley-2002-PRE,Dai-Xie-Jiang-Jiang-Zhou-2015-EmpE}.

\subsection*{Box clustering method}

The box clustering method has been applied to search for element clusters of the correlation matrix \cite{SalesPardo-Guimera-Moreira-Amaral-2007-PNAS}. It first determines the optimal ordering of matrix elements to ensure that the correlation matrix has a nested block-diagonal structure, where the simulated annealing approach is adopted to minimize the cost function:
\begin{equation}
  Q = \sum_{i,j=1}^{N} |i - j|C_{ij}
  \label{Eq: Box Clustering Cost Function}
\end{equation}
where $C_{ij}$ could be the element of raw or partial correlation matrix in this work. Then a greedy algorithm is implemented to partition time series into clusters. The procedures should be repeated $n$ times and obtain $n$ different partitions of HPI clusters \cite{Lancichinetti-Fortunato-2012-SR}. An affinity matrix $\mathbf{A}$ is obtained, whose element $A_{ij}$ is the number of partitions in which series $y_i(t)$ and $y_j(t)$ are assigned to the same cluster, divided by the number of partitions $n$. We take a typical number of $n = 1000$ here. Finally we apply the clustering method to the affinity matrix $\mathbf{A}$ itself, resulting in a final partition of the time series.

\subsection*{Partial correlations}

The concept of partial correlation is a powerful tool to investigate the intrinsic correlation between two time series effected by common factors \cite{Baba-Shibata-Sibuya-2004-ANZJS} and has been applied in stock markets \cite{Kenett-Shapira-BenJacob-2009-JPS,Kenett-Tumminello-Madi-GurGershgoren-Mantegna-BenJacob-2010-PLoS1,Kenett-Huang-Vodenska-Havlin-Stanley-2015-QF} and housing markets \cite{Meng-Xie-Jiang-Podobnik-Zhou-Stanley-2014-SR}. For time series $y_i(t)$, $i = 1,2,\ldots,N$ with a common collective trend $G(t)$, we can extract their idiosyncratic components $\varepsilon_i(t)$ by calibrating the following simple univariate factor model:
\begin{equation}
  y_i(t) = \alpha_i + \beta_iG(t) + \varepsilon_i(t).
\end{equation}
When there are more than one common factors, the above regression can be easily extended to the multivariate form. The correlation matrix of $\varepsilon_i(t)$ is the partial correlation matrix $\mathbf{P}$ of the original time series, whose elements $P_{ij}$ depict the residual correlations between $y_i(t)$ and $y_j(t)$ after removing the impact of the market-wide collective effect $G(t)$ which is the eigenportfolio of the largest eigenvalue.

\subsection*{Decomposition of correlation matrix}

With the complete set of eigenvalues eigenvalues and eigenvectors, the correlation matrix $\mathbf{C}$ can be expressed as follows:
\begin{equation}
\mathbf{C} = \sum_{i=1}^{N}v_i\lambda_iv'_i.
\end{equation}
Then, we can decompose the correlation matrix into three parts as \cite{Noh-2000-PRE,Kim-Jeong-2005-PRE}
\begin{equation}
\mathbf{C} = \mathbf{C}_m + \mathbf{C}_g + \mathbf{C}_r = v_1\lambda_1v'_1 + \sum_{i=2}^{N_g}v_i\lambda_iv'_i + \sum_{j=N_g+1}^{N}v_j\lambda_jv'_j,
\end{equation}
where the first component $\mathbf{C}_m$ represents a market mode reflecting collective behavior driven by a common influencing force, the second component $\mathbf{C}_g$ stands for the correlation structure with the bulk eigenvalues reflecting the partitioning of time series, and the third component $\mathbf{C}_r$ is the random noise terms. The determine of the market mode $\mathbf{C}_m$ is trivial. The determine of $N_g$ for $\mathbf{C}_g$ is not straightforward. One can use the eigenvalues that deviating the prediction of the RMT \cite{Plerou-Gopikrishnan-Rosenow-Amaral-Guhr-Stanley-2002-PRE} or estimate through econometric method \cite{Bai-Ng-2002-Em}. Because we have only 10 time series, we do not distinguish $\mathbf{C}_g$ and $\mathbf{C}_r$ and adopt the following simple decomposition:
\begin{equation}
  \mathbf{C} = \mathbf{C}_m + \mathbf{C}_b,
\end{equation}
where
\begin{equation}
  \mathbf{C}_m = v_1\lambda_1v'_1 ~~{\mathrm{and}} ~~ \mathbf{C}_b = \sum_{i=2}^{N}v_i\lambda_iv'_i.
\end{equation}
Note that, when $N$ is large, it is necessary to further extract the noise part $\mathbf{C}_r$.

\subsection*{The $\log t$ test}

The $\log t$ test proposed by Phillips and Sul is based on a nonlinear time varying factor model and provides a framework for modeling the transitional dynamics as well as long-run behaviors \cite{Phillips-Sul-2007-Em}. For the time series $y_i(t)$, we can represent it with a time varying common factor:
\begin{equation}
  y_i(t)=\delta_i(t)\mu(t),
  \label{Eq: yit decomposition}
\end{equation}
where $\mu(t)$ is a single common component and $\delta_i(t)$ is a time varying idiosyncratic element which captures the deviation of $i$ from the common path defined by $\mu(t)$. Following the previous work \cite{Phillips-Sul-2007-Em}, we eliminate the cyclical components by applying the HP filter \cite{Hodrick-Prescott-1997-JMCB} and extract the trend components $y_{\mathrm{hp},i}(t)$ of $y_i(t)$ as the analyzing series. Within this framework, all $N$ time series will converge, at some point in the future, to the steady state if $\lim_{k \rightarrow \infty}\delta_{i}(t+k)=\delta$ for all $i = 1,2,\ldots,N$, irrespective of whether the series are near the steady state or in transition. It is important given that the paths to the steady state across the time series can be significantly different. Since $\delta_i(t)$ cannot be directly estimated from Eq.~(\ref{Eq: yit decomposition}), Phillips and Sul eliminate the common component $\mu(t)$ through rescaling by the panel average \cite{Phillips-Sul-2007-Em}:
\begin{equation}
  h_i(t) = \frac{y_i(t)}{\frac{1}{N}\sum^{N}_{j=1}y_j(t)} = \frac{\delta_i(t)}{\frac{1}{N}\sum^{N}_{j=1}\delta_j(t)}.
  \label{Eq: h_it}
\end{equation}
The relative transition measurement $h_i(t)$ captures the transition path with respect to the panel average, which is analogical with the differential $D_i(t)$ in Eq.~(\ref{Eq: Deviation from Benchmark}). In order to define a formal econometric test of convergence as well as an empirical algorithm of defining club convergence, the following semi-parametric form for the time varying coefficients $\delta_i(t)$ is assumed:
\begin{equation}
  \delta_i(t) = \delta_i+\sigma_i(t)\xi_i(t),
  \label{Eq: delta_it}
\end{equation}
where $\sigma_i(t)=\frac{\sigma_i}{L(t)t^\alpha}, \sigma > 0, t \geq 0$, and $\xi_i(t)$ is weakly dependent upon $t$ but is $iid(0,1)$ over $i$.

The function $L(t)$ is a slow varying function, increasing and divergent at infinity ($L(t)=\ln t$ in the present report). Under this specific form for $\delta_i(t)$, the null hypothesis ${\mathcal{H}}_0$ of convergence for all $i$ and the alternative hypothesis ${\mathcal{H}}_1$ of non-convergence for some $i$ are expressed as follows:
\begin{equation}
  \left\{
  \begin{array}{lll}
    {\mathcal{H}}_0: \delta_i = \delta ~~{\mathrm{and}}~~ \alpha \geq 0  \\
    {\mathcal{H}}_1: \delta_i \neq \delta ~~{\mathrm{or}}~~ \alpha < 0
  \end{array}
  \right..
\end{equation}
Phillips and Sul demonstrate that the null of convergence can be tested in the framework of the following regression \cite{Phillips-Sul-2007-Em}:
\begin{equation}
  \ln({H_1}/{H_t})-2\ln L(t) = \hat{c}+\hat{b}\ln t+\hat{u_t}
  \label{Eq: regression model}
\end{equation}
for $t = [rT], [rT]+1,\ldots, T$, where $0 < r < 1$ (we use $r = 0.3$ in the present work as recommended in Ref.~\cite{Phillips-Sul-2007-Em}), $T$ is the length of initial time series, and $[rT]$ represents the integer part of $rT$. In this regression, $H_t = \frac{1}{N}\sum^{N}_{i=1}(h_{it}-1)^2$ and $\hat{b} = 2\hat{\alpha}$, where $h_{it}$ is relative transition path in Eq.~(\ref{Eq: h_it}) and $\hat\alpha$ is the least squares estimate of $\alpha$. The null hypothesis of convergence can be tested by applying a conventional one-side t-test for the slope coefficient $\hat{b} \geq 0$. For example, if the point estimate $\hat{b}$ is significantly less than zero, the null hypothesis of convergence is rejected. Specifically, at the $5\%$ significance level, the null hypothesis of convergence is rejected if the t statistic of $\hat{b}$ is less than -1.65, that is, $t_{\hat{b}} < -1.65$.

However, the rejection of full convergence does not imply the absence of convergence in subgroups of the panel. Phillips and Sul propose the following clustering algorithm to find a core convergence subgroup \cite{Phillips-Sul-2007-Em}:
\begin{enumerate}
\item Order the cities in the panel according to the last observation
\item Find core cities in the panel by running the $\log t$ regression for the $k$ highest cities with $2 \leq k \leq N$, and calculate the convergence $t$--statistic $t_k$. The core cities size is chosen on the basis of the maximum $t_k$ with $t_k > -1.65$.
\item Add one city at a time to the $k$ core member (step 2) and perform the $\log t$ test. If the resulting $t_k$ is greater than zero, a first convergence club is constituted.
\item Run a $\log t$ regression for the remaining cities in the panel and check if the convergence criterion is met. If this group satisfies the convergence test, then these members form a second convergence club. Otherwise, repeat step 1 to 3 to see if the remaining set can be further subdivided into convergence clusters. If no core group can be formed in, then these cities exhibit a divergent behavior.
\end{enumerate}

\section*{Results}

\subsection*{Raw correlation matrix of the raw HPI series}

\begin{figure}[htbp]
\centering
  \includegraphics[width=0.90\linewidth]{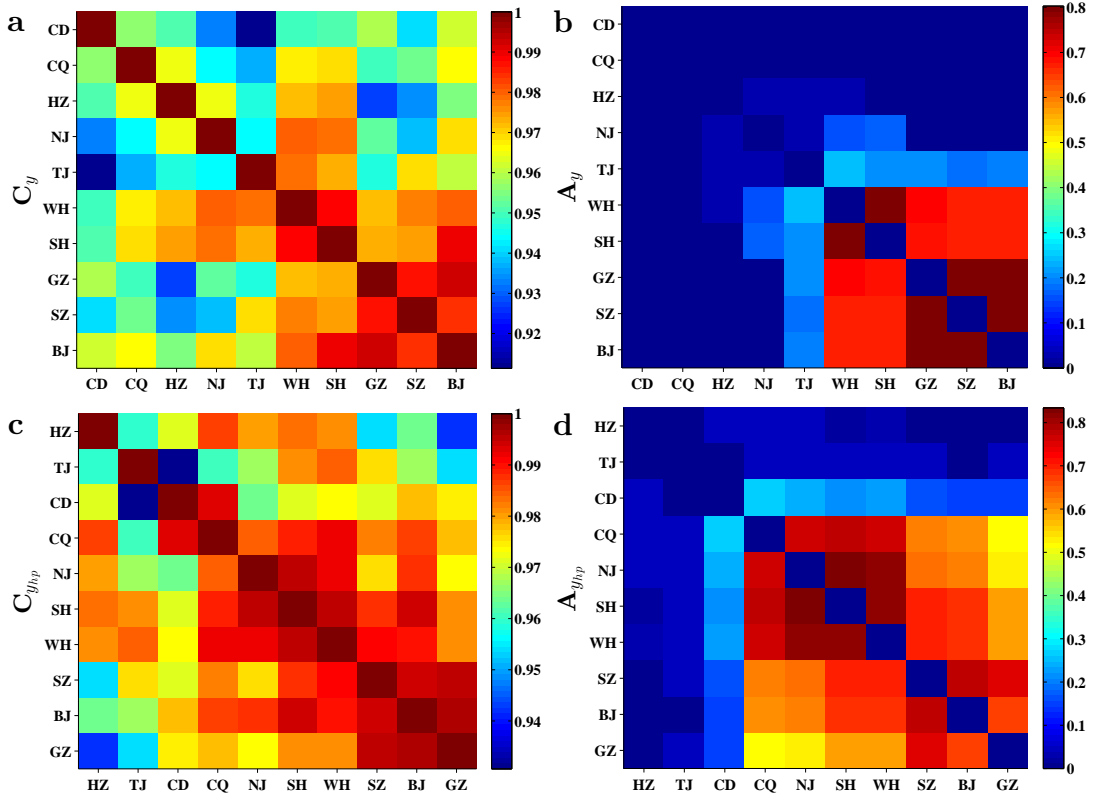}
  \caption{\label{Fig:CorrMat:AffiMat:Raw} {\textbf{Box clustering analysis of correlation matrices.}} (a) Correlation matrix of the 10 cities' initial HPI series $y_i(t)$ ordered according to the box clustering method. (b) Corresponding affinity matrix of the raw matrix. (c) Correlation matrix of $y_{{\mathrm{hp}},i}(t)$ ordered by box clustering method, which is the the trend component of $y_i(t)$ after eliminating the cyclical components by way of HP filter. (d) Corresponding affinity matrix of $y_{{\mathrm{hp}},i}(t)$.}
\end{figure}

The correlation matrix $\mathbf{C}_y$, whose element $C_{ij}$ is the Pearson correlation coefficient between the HPI series $y_i(t)$ of cities $i$ and $j$ has been studied. In the meanwhile, we also investigate the correlation matrix $\mathbf{C}_{y_{\rm{hp}}}$ of the trend component of $y_i(t)$ series, where the trend component $y_{\rm{hp}}$ is obtained by eliminating the cyclical component from $y_i(t)$ using HP filter \cite{Hodrick-Prescott-1997-JMCB}. Figure~\ref{Fig:CorrMat:AffiMat:Raw} illustrates the correlation matrices of $\mathbf{C}_y$ and $\mathbf{C}_{y_{\rm{hp}}}$ along with their corresponding affinity matrices $\mathbf{A}_y$ and $\mathbf{A}_{y_{\rm{hp}}}$ obtained by the box clustering method \cite{SalesPardo-Guimera-Moreira-Amaral-2007-PNAS}. Although box clustering method provides some block clusters in the affinity matrix, the extremely high correlation coefficients in the correlation matrix make the cluster results far-fetched. Nevertheless, we can still observe two clusters of cities in Fig.~\ref{Fig:CorrMat:AffiMat:Raw}{\textbf{b}}, in which Beijing, Guangzhou, Shenzhen and Shanghai are widely recognized as the first-tier cities in the Chinese housing market.

However, the eigenvalues and eigenvectors of $\mathbf{C}_y$ afford us some important information. Observing $\lambda_1$ and $v_1$ in Table \ref{TB:CorrMat:Eigen:Values:Vectors}, we can find that $v_1$ contains practically identical components with the same signs (positive here) and the contribution percent of $\lambda_1$ has reached an extremely high level with $\varphi_1 = 96.5\%$. According to the Random Matrix Theory \cite{Plerou-Gopikrishnan-Rosenow-Amaral-Guhr-Stanley-2002-PRE}, we conclude that there is a strong collective force driving these HPI series rising. The largest eigenvalue $\lambda_1$ and its eigenvector $v_1$ could adequately quantify the qualitative notion of the collective response of the entire system to stimuli. However, the partitioning effect of other large eigenvalues observed in stock markets is not evident for the 10 key cities.

\begin{table}[!ht]
\caption{{\textbf{Eigenvalues and eigenvectors of the raw correlation matrix.}} $\varphi_i$ is the percent of variability explained by the corresponding $\lambda_i$. $\phi_i$ is the cumulative percent of variability explained by $\lambda_1, \lambda_2, \cdots, \lambda_i$.}
\begin{tabular}{c|c c c c c c c c c c}\hline
& $v_1$ & $v_2$ & $v_3$ & $v_4$ & $v_5$ & $v_6$ & $v_7$ & $v_8$ & $v_9$ & $v_{10}$ \\\hline
Chengdu & 0.312 & 0.318 & -0.609 & -0.005 & -0.549 & 0.206 & 0.146 & -0.18 & 0.155 & 0.1 \\
Chongqing & 0.315 & 0.285 & -0.181 & -0.481 & 0.661 & 0.254 & 0.175 & 0.111 & 0.073 & -0.074 \\
Hangzhou & 0.314 & 0.538 & 0.231 & -0.15 & -0.201 & -0.473 & -0.376 & 0.312 & -0.119 & -0.137 \\
Nanjing & 0.315 & 0.208 & 0.304 & 0.674 & 0.143 & 0.289 & 0.03 & 0.152 & 0.428 & 0.001 \\
Tianjin & 0.313 & -0.305 & 0.479 & -0.381 & -0.403 & 0.202 & 0.417 & 0.22 & 0.067 & -0.062 \\
Wuhan & 0.32 & -0.017 & 0.21 & 0.017 & -0.002 & 0.46 & -0.398 & -0.345 & -0.563 & 0.214 \\
Shanghai & 0.32 & 0.003 & 0.162 & 0.073 & 0.121 & -0.434 & 0.309 & -0.716 & 0.02 & -0.232 \\
Guangzhou & 0.317 & -0.376 & -0.342 & 0.255 & 0.034 & -0.003 & -0.043 & 0.274 & -0.31 & -0.634 \\
Shenzhen & 0.317 & -0.468 & -0.11 & -0.199 & 0.045 & -0.136 & -0.542 & -0.086 & 0.531 & 0.158 \\
Beijing & 0.32 & -0.176 & -0.149 & 0.191 & 0.142 & -0.355 & 0.285 & 0.269 & -0.269 & 0.66 \\\hline
$\lambda_i$ & 9.65 & 0.113 & 0.104 & 0.058 & 0.036 & 0.018 & 0.012 & 0.005 & 0.002 & 0.001 \\\hline
$\varphi_i$ & 96.5\% & 1.13\% & 1\% & 0.58\% & 0.36\% & 0.18\% & 0.12\% & 0.05\% & 0.02\% & 0.01\% \\\hline
$\phi_i$ & 96.5\% & 97.63\% & 98.67\% & 99.25\% & 99.62\% & 99.8\% & 99.92\% & 99.97\% & 99.99\% & 100\% \\\hline
\end{tabular}
\label{TB:CorrMat:Eigen:Values:Vectors}
\end{table}

\subsection*{Partial correlation of the raw HPI series}

According to Table \ref{TB:CorrMat:Eigen:Values:Vectors}, the projection on $v_1$ would contain $96.5\%$ of the total variability of the 10 cities' HPIs. Thus we compute the eigenportfolio associated with $\lambda_1$ as follows:
\begin{equation}
G(t) = u_1^{\mathbf{T}}\mathbf{y(t)},
\label{Eq: Eigenportfolio}
\end{equation}
which can be treated as the benchmark of the collective rising trend. $u_1^{\mathbf{T}}$ is a $1 \times 10$ vector whose components are the square components of $v_1$ and $\mathbf{y(t)}$ is a $10 \times 101$ matrix which contains the original HPI series of the 10 cities. The square and normalization procedure from $v_1$ to $u_1$ is to make sure that the sum of its components is 1 and $G(t)$ has identical magnitude with the initial HPI series.

\begin{figure}[htbp]
\centering
  \includegraphics[width=0.9\linewidth]{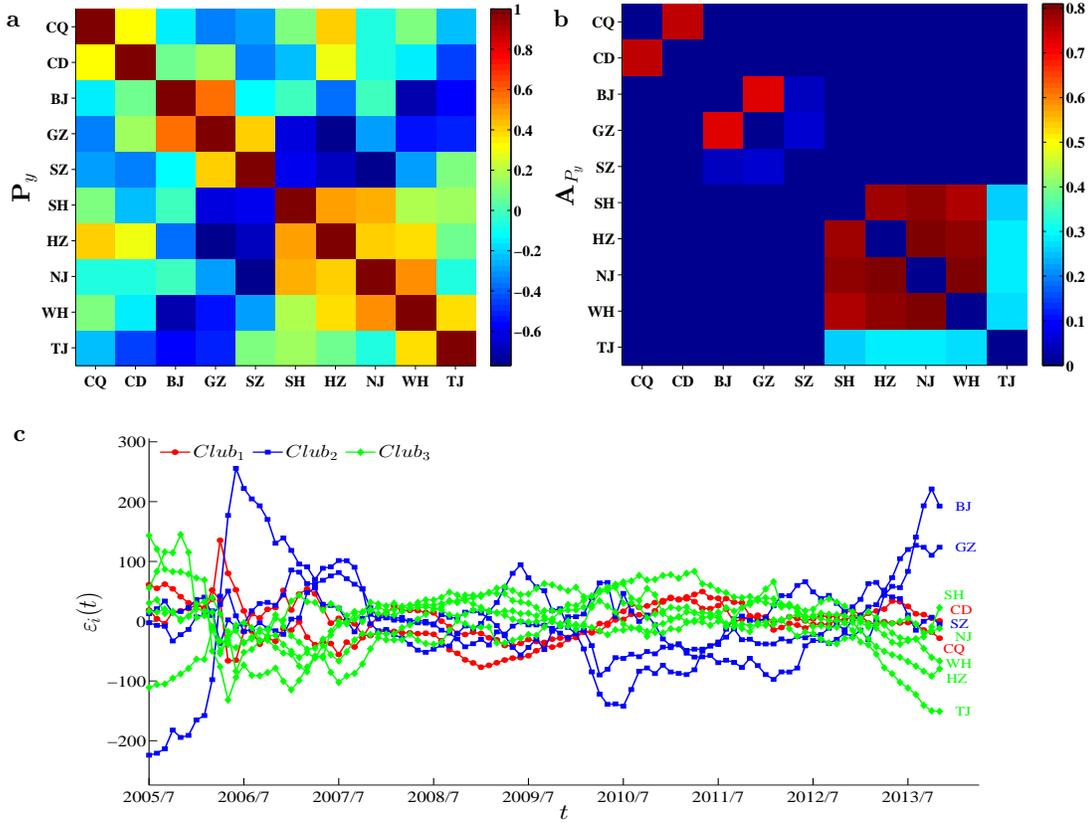}
  \caption{\label{Fig:PartialCorr:Mat} {\bf{Partial correlation analysis.}} (a) The partial correlation matrix of initial HPI series. (b) The corresponding affinity matrix obtained by the box clustering method. (c) Residual series $\varepsilon_i(t)$ obtained by regressing $y_i(t)$ in respect to the collective trend $G(t)$. Varying colors and marks represent block clubs obtained by box clustering method.}
\end{figure}

With the eigenportfolio $G(t)$ acting as the collective trend, we can calculate the partial correlation matrix $\mathbf{P}_y$, whose elements $P_{ij}$ are the partial correlation coefficients of cities $i$ and $j$. Figure~\ref{Fig:PartialCorr:Mat} demonstrates $\mathbf{P}_y$ and its corresponding affinity matrix $\mathbf{A}_{\mathbf{P}_y}$. One can observe that, the correlation relationship between the $y_i(t)$ weakens after eliminating the impact of $G(t)$ in the sense that the elements of $\mathbf{P}_y$ are significantly smaller than those of $\mathbf{C}_y$.

According to the affinity matrix $\mathbf{A}_{\mathbf{P}_y}$ in Fig.~\ref{Fig:PartialCorr:Mat}{\textbf{b}}, we observe three clusters of cities: Club$_1$: Chongqing (CQ) and Chengdu (CD); Club$_2$: Beijing (BJ), Guangzhou (GZ) and Shenzhen (SZ); and Club$_3$: Shanghai (SH), Nanjing (NJ), Wuhan (WH), Hangzhou (HZ) and Tianjin (TJ). The residual series $\varepsilon_i(t)$ and their corresponding clubs are demonstrated in Fig.\ref{Fig:PartialCorr:Mat}{\textbf{c}}. It is evident that the trajectories in the same club have a similar pattern, while the paths in different clubs exhibit different shape.

\subsection*{Decomposition of correlation matrix of the raw HPI series}

\begin{figure}[ht!]
  \centering
  \includegraphics[width=0.98\linewidth]{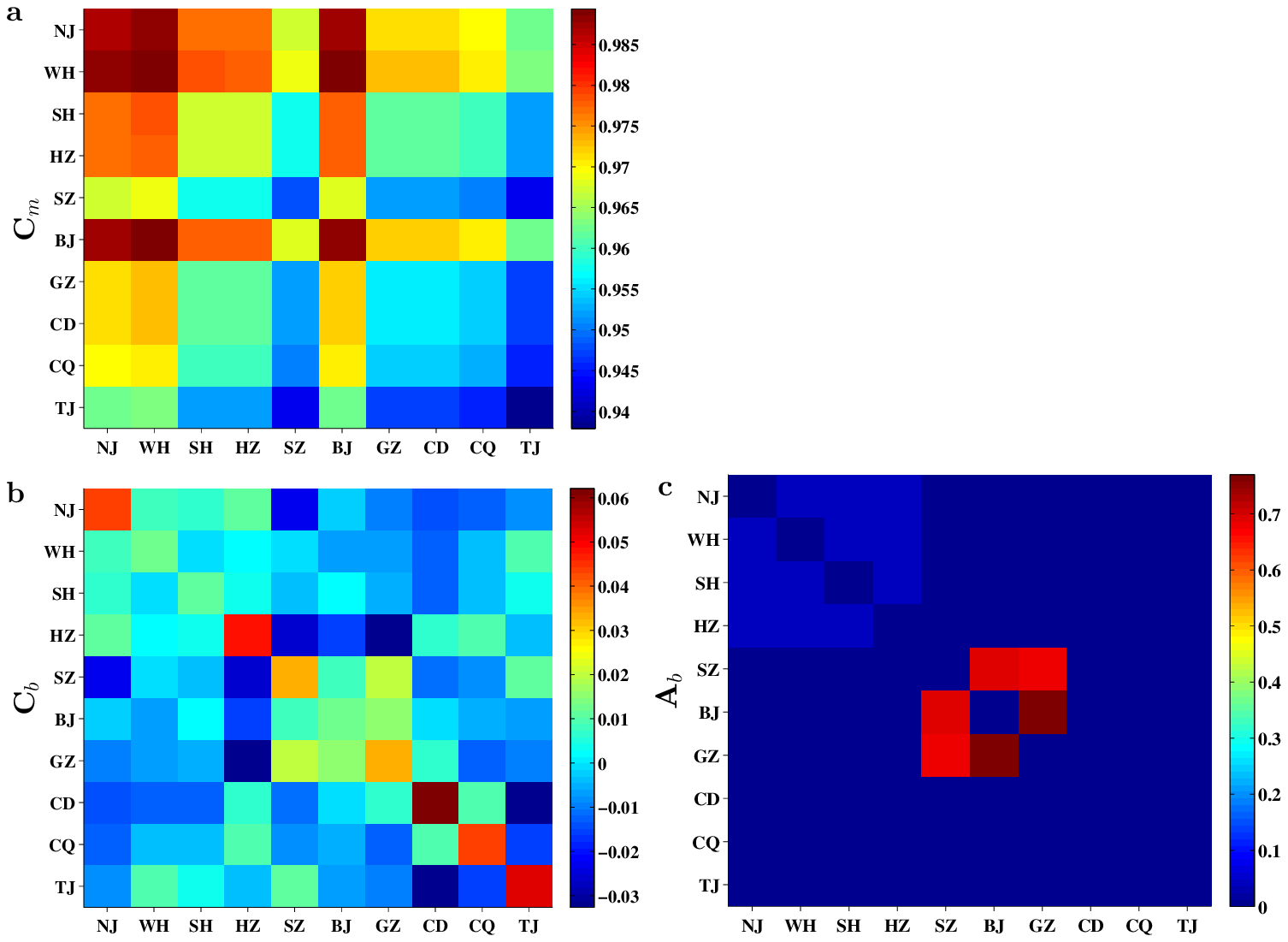}
  \vspace{2mm}
  \caption{\label{Fig:Decomposition:Corr:Affinity:Matrix} {\textbf{Decomposition of the raw correlation matrix.}} (a) Matrix $\mathbf{C}_m$ reflecting the market effect. (b)Residual matrix $\mathbf{C}_b$. (c) Affinity matrix $\mathbf{A}_b$ of the residual matrix $\mathbf{C}_b$.}
\end{figure}

We decompose the raw matrix correlation matrix $\mathbf{C}_y$ into the market effect part $\mathbf{C}_m$ and the residual part $\mathbf{C}_b$, as illustrated in  Fig.~\ref{Fig:Decomposition:Corr:Affinity:Matrix}{\textbf{a}} and {\textbf{b}}. We can see that elements of $\mathbf{C}_b$ are much smaller than the corresponding elements in $\mathbf{C}_m$. This suggests that the majority of the extremely large correlation coefficients of $\mathbf{C}_y$ come from the marketwide collective trend. For the raw HPI series $y_i(t)$, little information is left after removing the strong collective trend.

Figure~\ref{Fig:Decomposition:Corr:Affinity:Matrix}{\textbf{c}} shows the affinity matrix $\mathbf{A}_b$ by implementing the box clustering method on the residual matrix $\mathbf{C}_b$. The cluster in the center of $\mathbf{A}_b$ contains SZ, GZ and BJ, which is the most evident. At the northwest corner of $\mathbf{A}_b$, we see another cluster containing NJ, WH, SH and HZ. The cluster for CQ, CD and TJ is not clear. Therefore, the decomposed components of the raw matrix are also able to categorize city clusters with similar evolution of the house price indexes.

\subsection*{Analysis on differentials between HPI series $y_i(t)$ and the collective trend}

\begin{figure}[htbp]
  \centering
  \includegraphics[width=0.90\linewidth]{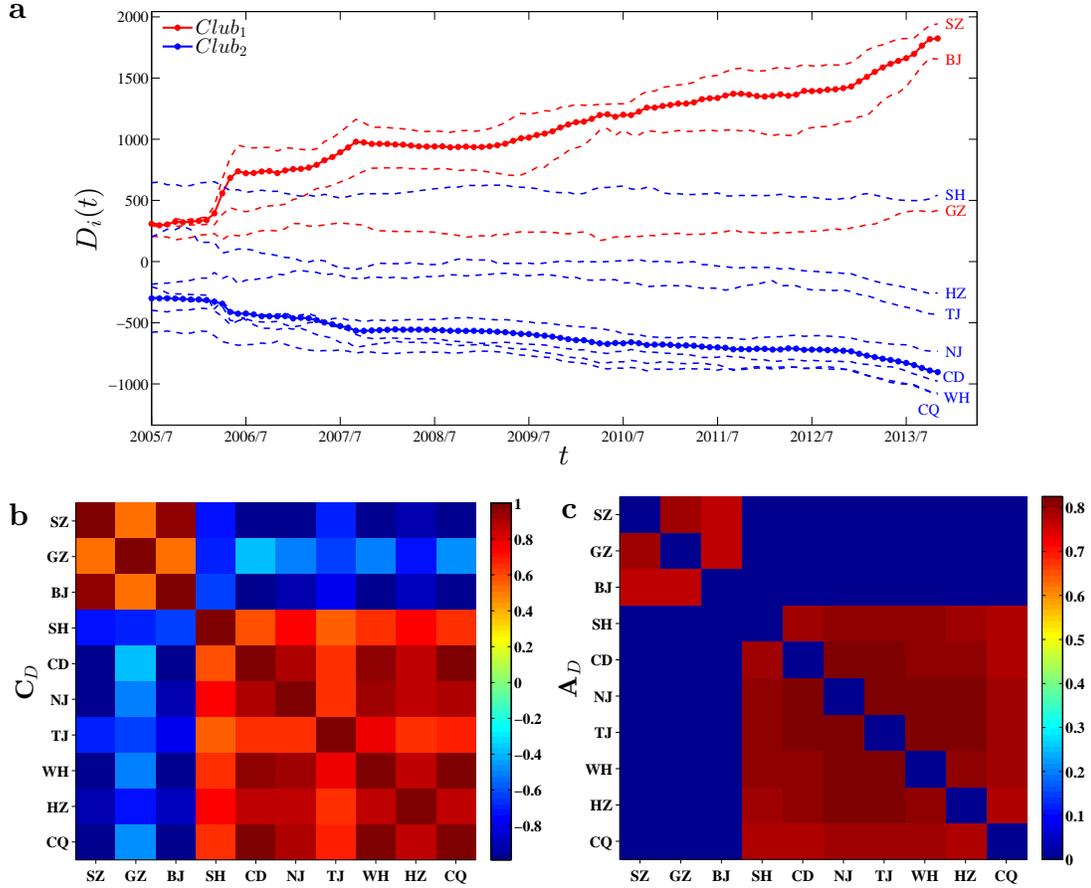}
  \caption{\label{Fig:Dy:Clubs} {\textbf{Analyzing the differentials between HPI series $y_i(t)$ and the collective trend.}} (a) Evolution of the differentials between the HPI time series and the collective trend. The three red dashed lines correspond to Beijing, Shenzhen and Guangzhou in Club$_1$, while the blue dashed lines correspond to the seven remaining cities in Club$_2$. The two continuous lines decorated with solid circles highlight the individual collective trends $G_{{\mathrm{Club}}_1}(t)$ and $G_{{\mathrm{Club}}_2}(t)$ of the two clubs respectively. (b) The correlation matrix $\mathbf{C}_D$ determined by the deviation paths of the 10 cities. (c) Corresponding affinity matrix $\mathbf{A}_D$ obtained by box clustering method.}
\end{figure}

We now turn to investigate the relative behavior of the HPI to their collective trends. The deviation from the collective trend can be quantified by the differential between regional HPI $y_i(t)$ and collective trend benchmark $G(t)$:
\begin{equation}
  D_i(t)=y_i(t)-G(t).
  \label{Eq: Deviation from Benchmark}
\end{equation}
In Fig.~\ref{Fig:Dy:Clubs}{\textbf{a}}, we show the deviation paths $D_i(t)$ of the ten cities. One can intuitively observe that the ten paths fall into two groups according to their trends: rising up and falling down. The rising-up group contains Shenzhen, Beijing and Guangzhou, showing high-than-average house price growth.

Fig.~\ref{Fig:Dy:Clubs}{\textbf{b}} shows the cross-correlation matrix $\mathbf{C}_{D}$ of the deviation paths. There are two obvious blocks with positive correlations within the blocks and negative correlations between the blocks, consistent with the opposite trends in the differentials in Fig.~\ref{Fig:Dy:Clubs}{\textbf{a}}. The collective effect has been successfully removed. We also apply the box clustering method \cite{SalesPardo-Guimera-Moreira-Amaral-2007-PNAS} to the correlation matrix $\mathbf{C}_{D}$. By sorting the cities according to the orders of affinity matrix $A$, two significant clubs are visualized in Fig.~\ref{Fig:Dy:Clubs}{\textbf{c}}. Club$_1$ includes three cites Shenzhen, Guangzhou and Beijing, and Club$_2$ includes seven cities Shanghai, Chengdu, Nanjing, Tianjin, Wuhan, Hangzhou and Chongqing.

The eigenvectors and eigenvalues of the differential matrix $\mathbf{C}_{D}$ are presented in Table~\ref{Tb:Eigenvalue:Eigenvector:C:Dy}. We find that the components of the largest eigenvector $v_1$ have positive and negative signs, corresponding respectively to the two clubs identified in Fig.~\ref{Fig:Dy:Clubs}. It is not surprising that the largest eigenvector does not reflect the market effect any longer but includes some grouping information, which is reminiscent of the US housing market \cite{Meng-Xie-Jiang-Podobnik-Zhou-Stanley-2014-SR}. The nine remaining eigenvectors do not possess much economic information. Note that $\lambda_1$ is much larger than other $\lambda_i$'s and the large value of $\phi_1$ also indicates the high systemic risk in the Chinese housing market.

\begin{table}[htbp]
\centering
  \caption{{\textbf{Eigenvectors and eigenvalues of the differential matrix $\mathbf{C}_{D}$.}} The eigenvalues $\lambda_i$ and the corresponding eigenvectors $v_i$ of the correlation matrix $\mathbf{C}_{D}$ of deviation paths $D_i(t)$, $i=1,2, \cdots, 10$. $\varphi_i$ is the percent of variability explained by the corresponding $\lambda_i$. $\phi_i$ is the cumulative percents of variability explained by $\lambda_1, \lambda_2, \cdots, \lambda_i$.}
\begin{tabular}{cc c c c c c c c c c}\hline
& $v_1$ & $v_2$ & $v_3$ & $v_4$ & $v_5$ & $v_6$ & $v_7$ & $v_8$ & $v_9$ & $v_{10}$ \\\hline
Shenzhen & 0.346 & 0.149 & 0.139 & 0.046 & -0.056 & -0.278 & 0.463 & -0.077 & -0.338 & 0.649 \\
Guangzhou & 0.227 & -0.732 & -0.082 & -0.421 & -0.069 & 0.319 & 0.241 & 0.238 & 0.009 & 0.087 \\
Beijing & 0.343 & 0.172 & -0.169 & -0.036 & 0.143 & 0.274 & -0.504 & 0.051 & 0.429 & 0.536 \\
Shanghai & -0.271 & 0.416 & -0.488 & -0.628 & -0.308 & 0.005 & 0.15 & -0.012 & 0 & 0.065 \\
Chengdu & -0.332 & -0.307 & -0.034 & 0.129 & -0.317 & 0.205 & -0.152 & -0.724 & -0.085 & 0.284 \\
Nanjing & -0.332 & -0.113 & -0.216 & -0.158 & 0.8 & -0.049 & -0.085 & -0.052 & -0.346 & 0.183 \\
Tianjin & -0.273 & 0.203 & 0.783 & -0.369 & 0.027 & 0.326 & -0.056 & 0.084 & -0.039 & 0.129 \\
Wuhan & -0.345 & -0.162 & 0.102 & -0.011 & 0.142 & -0.382 & 0.33 & -0.031 & 0.73 & 0.19 \\
Hangzhou & -0.331 & 0.136 & -0.19 & 0.49 & -0.013 & 0.559 & 0.371 & 0.342 & 0.006 & 0.169 \\
Chongqing & -0.34 & -0.216 & 0 & 0.088 & -0.339 & -0.369 & -0.415 & 0.531 & -0.2 & 0.291 \\\hline
$\lambda_i$ & 8.016 & 0.984 & 0.521 & 0.273 & 0.112 & 0.054 & 0.023 & 0.011 & 0.006 & 0 \\
$\varphi_i$ & 80.16\% & 9.84\% & 5.21\% & 2.73\% & 1.12\% & 0.54\% & 0.23\% & 0.11\% & 0.06\% & 0\% \\
$\phi_i$ & 80.16\% & 90\% & 95.21\% & 97.94\% & 99.06\% & 99.6\% & 99.83\% & 99.94\% & 100\% & 100\% \\\hline
\end{tabular}
\label{Tb:Eigenvalue:Eigenvector:C:Dy}
\end{table}

We further study the two correlation matrices of the differentials in Club$_1$ and Club$_2$, resulting in a $3\times3$ matrix for Club$_1$ and a $7\times7$ matrix for Club$_2$. The eigenvalues and eigenvectors are listed in the Table~\ref{Tab:Eigen:Value:Vector:Club:Dy}. It is found that the components of the two eigenvectors $v_1$ associated with the two largest eigenvalues $\lambda_1$ have same signs and relatively similar magnitudes. Hence, these eigenvectors indicate the presence of a collective behavior within the two subsystems \cite{Meng-Xie-Jiang-Podobnik-Zhou-Stanley-2014-SR}. It is interesting to notice that the relative magnitudes of $v_1$ components for the two clubs are similar to the whole matrix in Table \ref{TB:CorrMat:Eigen:Values:Vectors}. For instance, the results show that $v_{1,{\mathrm{Shenzhen}}}= 0.622$, $v_{1,{\mathrm{Guangzhou}}}= 0.474$ and $v_{1,{\mathrm{Beijing}}}=0.623$ for Club$_1$ (Table \ref{Tb:Eigenvalue:Eigenvector:C:Dy}) and $v_{1,{\mathrm{Shenzhen}}}= 0.346$, $v_{1,{\mathrm{Guangzhou}}}= 0.227$ and $v_{1,{\mathrm{Beijing}}}=0.343$ for the whole matrix (Table \ref{TB:CorrMat:Eigen:Values:Vectors}). In both cases, we have $v_{1,{\mathrm{Shenzhen}}}:v_{1,{\mathrm{Guangzhou}}}:v_{1,{\mathrm{Beijing}}}\approx3:2:3$. It implies that Shenzhen and Beijing dominate the collective behavior in Club$_1$, while the contribution of Guangzhou is relatively smaller. For Club$_2$, Shanghai and Tianjin have relative small $v_1$ components, while other components are close to each other. We determine the eigenportfolios of both clubs according to Eq.~(\ref{Eq: Eigenportfolio}) to extract their common trends $G_{{\mathrm{Club}}_1}(t)$ and $G_{{\mathrm{Club}}_2}(t)$, which are demonstrated in Fig.~\ref{Fig:Dy:Clubs}{\textbf{a}}. Obviously, $G_{{\mathrm{Club}}_1}(t)$ has an increasing trend and $G_{{\mathrm{Club}}_2}(t)$ has a decreasing trend. It does not mean that the house prices in Club$_1$ rise up while the house prices in Club$_2$ fall down. Instead, it means that the house prices in Club$_1$ grow faster than average while the house prices in Club$_2$ grows slower than average. Different from the case of the UK housing market \cite{Holmes-Grimes-2008-US}, the eigenportfolios of the two clubs are non-stationary and present remarkable trends over time $t$.

\begin{table}[htbp]
\centering
\caption{{\textbf{The eigenvalues $\lambda_i$ and the corresponding eigenvectors $v_i$ of the cross-correlation matrix of deviation paths of Club$_1$ and Club$_2$ respectively.}} The two clubs are obtained from the box clustering method shown in Fig.~\ref{Fig:Dy:Clubs}{\textbf{c}}. $\varphi_i$ is the percents of variability explained by the corresponding $\lambda_i$. $\phi_i$ is the cumulative percents of variability explained by $\lambda_1,\lambda_2, \cdots \lambda_i$.}
\begin{tabular}{ccc c c c c c c}\hline
\centering
  Club$_1$& $v_1$ & $v_2$ & $v_3$ \\\cline{1-4}
  Shenzhen & 0.622 & 0.339 & 0.706 \\
  Guangzhou & 0.474 & -0.88 & 0.004 \\
  Beijing & 0.623 & 0.332 & -0.708 \\\cline{1-4}
  $\lambda_i$ & 2.34 & 0.611 & 0.049 \\
  $\varphi_i$ & 78\% & 20.4\% & 1.6\% \\
  $\phi_i$ & 78\% & 98.4\% & 100\% \\\hline
  Club$_2$& $v_1$ & $v_2$ & $v_3$ & $v_4$ & $v_5$ & $v_6$ & $v_7$ \\\hline
  Shanghai & 0.32 & 0.864 & -0.136 & 0.172 & -0.307 & -0.029 & -0.081 \\
  Chengdu & 0.396 & -0.284 & 0.275 & 0.011 & -0.382 & -0.709 & -0.198 \\
  Nanjing & 0.396 & 0.06 & 0.198 & 0.458 & 0.693 & -0.172 & 0.283 \\
  Tianjin & 0.32 & -0.228 & -0.9 & -0.049 & 0.048 & -0.136 & 0.11 \\
  Wuhan & 0.41 & -0.214 & 0.031 & 0.194 & 0.059 & 0.495 & -0.707 \\
  Hangzhou & 0.388 & 0.154 & 0.154 & -0.847 & 0.288 & 0.045 & 0.009 \\
  Chongqing & 0.404 & -0.217 & 0.178 & 0.06 & -0.436 & 0.449 & 0.602 \\\cline{1-8}
  $\lambda_i$ & 5.712 & 0.524 & 0.477 & 0.148 & 0.106 & 0.023 & 0.011 \\
  $\varphi_i$ & 81.6\% & 7.5\% & 6.8\% & 2.1\% & 1.5\% & 0.3\% & 0.2\% \\
  $\phi_i$ & 81.6\% & 89.1\% & 95.9\% & 98\% & 99.5\% & 99.8\% & 100\% \\\hline
\end{tabular}
\label{Tab:Eigen:Value:Vector:Club:Dy}
\end{table}

The relative small percents of $\varphi_{1} = 78\%$ for Club$_1$ and $\varphi_{1} = 81.6\%$ for Club$_2$ implies that, there are still remarkable portions of variabilities hidden in the rest of eigenvalues and eigenvectors after extracting the common trend of $G_{{\mathrm{Club}}_1}(t)$ and $G_{{\mathrm{Club}}_2}(t)$. Scrutinizing the contents of eigenvectors of Club$_1$, we already notice that the loading of Guangzhou on $G_{{\mathrm{Club}}_1}(t)$ with $v_{1,{\mathrm{Guangzhou}}} = 0.474$ is relative smaller than the other two cites with $v_{1,{\mathrm{Shenzhen}}} = 0.622$ and $v_{1,{\mathrm{Beijing}}} = 0.623$, which indicates that Guangzhou, as a component of $G_{{\mathrm{Club}}_1}(t)$, doesn't contribute to the club's collective tendency as significantly as Shenzhen and Beijing do. Conversely, the loading of $v_{2,{\mathrm{Guangzhou}}} = -0.88$ would lead $v_2$ to having a large inverse participation ratio (IPR), which is often applied in localization theory, suggesting that $v_2$ is localized due to the significant contribution of Guangzhou on it. Therefore eigenvector $v_2$ would include information of heterogeneity of Guangzhou in Club$_1$. The eigenvector $v_3$ also contains significant participation contents, namely the loadings $v_{3,{\mathrm{Shenzhen}}} = 0.706$ and $v_{3,{\mathrm{Beijing}}} = -0.708$, with relative negative signs. This pair of components in eigenvector $v_3$ associated with the smallest eigenvalue $\lambda_3$ highlights a considerable linear relationship between the two participants Shenzhen and Beijing, which has the largest correlation coefficient $C_{{\mathrm{Shenzhen, Beijing}}} = 0.9509$ in Club$_1$ \cite{Plerou-Gopikrishnan-Rosenow-Amaral-Guhr-Stanley-2002-PRE}. Investigating Club$_2$ in the same way, we find that cities like Shanghai, Tianjin and Hangzhou do not contribute to the collective tendency as significantly as other cities do, according to their relative small loadings on $v_1$. Their heterogeneities have dispersed in the rest eigenvectors with significantly ``large'' loadings like $v_{2,{\mathrm{Shanghai}}} = 0.864, v_{3,{\mathrm{Tianjin}}} = -0.9$ and $v_{4,{\mathrm{Hangzhou}}} = -0.847$. In addition, one can still observe less heterogeneity in the $v_5$ components $v_{5,{\mathrm{Nanjing}}} = 0.693$ and $v_{6,{\mathrm{Chengdu}}} = -0.709$. Similarly, pairs of components $v_{7,{\mathrm{Wuhan}}} = -0.707$ and $v_{7,{\mathrm{Chongqing}}} = 0.602$ highlights the strong linearity between the two cities with a large correlation coefficient $C_{{\mathrm{Wuhan, Chongqing}}} = 0.977$.

\subsection*{$\log t$ test analysis of convergence}

The blocks or clusters identified so far are obtained by different methods based on linear correlation coefficients. It is not unusual that there are nonlinear relationships between elements in complex economic systems. Therefore, we adopt an alternative econometric technique called the $\log t$ test to consolidate our results \cite{Phillips-Sul-2007-Em}. The null hypothesis of convergence can be tested by applying a conventional one-side t-test for the slope coefficient $\hat{b} \geq 0$. If the point estimate $\hat{b}$ is significantly less than zero, the null hypothesis of convergence is rejected. At the $5\%$ significance level, the critical value is $t_c = -1.65$.

Table \ref{Tab: log t Club} reports the results of the $\log t$ test. The null hypothesis of overall convergence of 10 cities' HPI is rejected at the $5\%$ significance level since the t-statistic $t_{\hat{b}} = -51.25$ is far less than the critical value $-1.65$. However, our analysis identifies four clubs: Club$_1$ (Beijing, Shenzhen), Club$_2$ (Shanghai, Guangzhou), Club$_3$ (Tianjin, Hangzhou), and Club$_4$ (Nanjing, Wuhan, Chengdu, Chongqing). All the $\hat{b}$ coefficients are positive and the t-statistics are larger than $t_c$. The identification of Club$_1$ with Shenzhen and Beijing is consistent with the results from linear methods. In addition, grouping Shanghai and Guangzhou as Club$_2$ meets our common perception of the Chinese housing market.

\begin{table}[htbp]
\centering
\caption{{\textbf{Club convergence obtained by the $\log t$ test.}} }
\begin{tabular}{c c c c c}\hline
& Club$_1$ & Club$_2$ & Club$_3$ & Club$_4$ \\\hline
\multirow{4}*{All Cities} & \multirow{3}*{Shenzhen} & \multirow{3}*{Shanghai} & \multirow{3}*{Hangzhou} & Nanjing\\
& \multirow{3}*{Beijing}   & \multirow{3}*{Guangzhou}    & \multirow{3}*{Tianjin}  & Wuhan \\
& & & & Chengdu\\
& & & & Chongqing\\\hline
$t_{\hat{b}} =-51.25$ & $t_{\hat{b}} = 0.60$ & $t_{\hat{b}} = -0.84$ & $t_{\hat{b}} = 2.61$ & $t_{\hat{b}} = 2.44$ \\[1mm]
   $\hat{b} = -0.86$  & $\hat{b} = 0.038$     & $\hat{b} = 0.060$      & $\hat{b} = 0.47$     & $\hat{b} = 0.095$ \\\hline
\end{tabular}
\label{Tab: log t Club}
\end{table}

\begin{figure}[htbp]
\centering
\includegraphics[width=13cm]{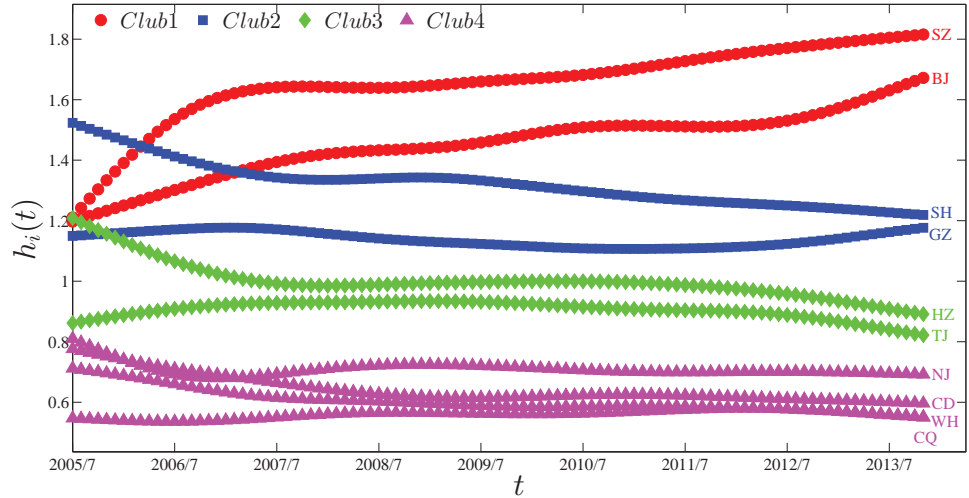}
\caption{{\textbf{The relative transitional paths $h_i(t)$ of the 10 cities.}} Different colors stand for different clubs obtained by the $\log t$ test.}
\label{Fig: Hy Series Club}
\end{figure}

In Fig.~\ref{Fig: Hy Series Club}, we illustrate the relative transitional paths $h_i(t)$ of the 10 cities. For each club, the evolution of the relative transitional paths $h_i(t)$ may be relatively irrelevant at the early stage. However, they exhibit a clear convergence in the latest years.

\section*{Conclusion and discussion}

In summary, we aimed at quantifying the behaviors of HPI series of 10 key cities of China, based on both linear and non-linear approaches. An extremely strong collective trend has been detected, driving all the HPI series rising. Simultaneously, according to the investigation of partial correlation and residual information matrix, it also shows that correlations between series basically come from this collective trend.

The relative behaviors of HPI series to their collective trend are also studied. Deviation paths, which is quantified by the differentials between HPI series and their collective trend, are grouped into two clubs, with Club$_1$ consisting of Shenzhen, Guangzhou and Beijing and Club$_2$ consisting of Shanghai, Chengdu, Nanjing, Tianjin, Wuhan, Hangzhou and Chongqing. Members between the two clubs are anti-correlated, corresponding to the deviation paths going towards opposite directions. It suggests that the rising of the Chinese HPI is driven by a minority of cities like those in Club$_1$. Some heterogeneities for the two clubs can be observed by investigating the eigenvalues and eigenvectors of the correlation matrix.

A recent panel convergence test, namely $\log t$ test, has been applied to examine the convergence of the HPI series. It reveals that 10 cities' HPI series do not form a homogeneous convergence club. Instead, our results identify four city clubs that converge to different steady states. In subsequent studies, it would be quite interesting to tackle the lead-lag structure of the HPIs to pinpoint the propagation mechanisms within the Chinese housing market.

\section*{Acknowledgments}

This work was partially supported by the National Natural Science Foundation of China (71131007), the Program for Changjiang Scholars and Innovative Research Team in University (IRT1028), the Shanghai (Follow-up) Rising Star Program (Grant No. 11QH1400800), and the Shanghai ``Chen Guang'' Project (2012CG34), and the Fundamental Research Funds for the Central Universities.


\begin{thebibliography}{10}

\bibitem{Sornette-Zhou-2004-PA}
Sornette D, Zhou WX.
\newblock {Evidence of fueling of the 2000 new economy bubble by foreign
  capital inflow: Implications for the future of the US economy and its stock
  market}.
\newblock Physica A. 2004;332:412--440.

\bibitem{Sornette-2003}
Sornette D.
\newblock {Why Stock Markets Crash}.
\newblock Princeton: Princeton University Press; 2003.

\bibitem{Sornette-2003-PR}
Sornette D.
\newblock {Critical market crashes}.
\newblock Phys Rep. 2003;378:1--98.

\bibitem{Zhou-Sornette-2003a-PA}
Zhou WX, Sornette D.
\newblock {2000-2003 real estate bubble in the UK but not in the USA}.
\newblock Physica A. 2003;329:249--263.

\bibitem{Zhou-Sornette-2006b-PA}
Zhou WX, Sornette D.
\newblock {Is there a real-estate bubble in the US?}
\newblock Physica A. 2006;361:297--308.

\bibitem{Jiang-Zhou-Sornette-Woodard-Bastiaensen-Cauwels-2010-JEBO}
Jiang ZQ, Zhou WX, Sornette D, Woodard R, Bastiaensen K, Cauwels P.
\newblock {Bubble diagnosis and prediction of the 2005-2007 and 2008-2009
  Chinese stock market bubbles}.
\newblock J Econ Behav Org. 2010;74:149--162.

\bibitem{Zhou-Sornette-2004a-PA}
Zhou WX, Sornette D.
\newblock {Antibubble and prediction of China's stock market and real-estate}.
\newblock Physica A. 2004;337:243--268.

\bibitem{Schweitzer-Fagiolo-Sornette-VegaRedondo-Vespignani-White-2009-Science}
Schweitzer F, Fagiolo G, Sornette D, Vega-Redondo F, Vespignani A, White DR.
\newblock {Economic networks: The new challenges}.
\newblock Science. 2009;325:422--425.

\bibitem{Kritzman-Li-Page-Rigobon-2011-JPM}
Kritzman M, Li YZ, Page S, Rigobon R.
\newblock {Principal components as a measure of systemic risk}.
\newblock J Portf Manag. 2011;37(4):112--126.

\bibitem{Billio-Getmansky-Lo-Pelizzon-2012-JFE}
Billio M, Getmansky M, Lo AW, Pelizzon L.
\newblock {Econometric measures of connectedness and systemic risk in the
  finance and insurance sectors}.
\newblock J Financial Econ. 2012;104(3):535--559.

\bibitem{Zheng-Podobnik-Feng-Li-2012-SR}
Zheng ZY, Podobnik B, Feng L, Li BW.
\newblock {Changes in cross-correlations as an indicator for systemic risk}.
\newblock Sci Rep. 2012;2:888.

\bibitem{Meng-Xie-Jiang-Podobnik-Zhou-Stanley-2014-SR}
Meng H, Xie WJ, Jiang ZQ, Podobnik B, Zhou WX, Stanley HE.
\newblock {Systemic risk and spatiotemporal dynamics of the US housing market}.
\newblock Sci Rep. 2014;4:3566.

\bibitem{Laloux-Cizean-Bouchaud-Potters-1999-PRL}
Laloux L, Cizeau P, Bouchaud JP, Potters M.
\newblock {Noise dressing of financial correlation matrices}.
\newblock Phys Rev Lett. 1999;83:1467--1470.

\bibitem{Plerou-Gopikrishnan-Rosenow-Amaral-Stanley-1999-PRL}
Plerou V, Gopikrishnan P, Rosenow B, Amaral LAN, Stanley HE.
\newblock {Universal and nonuniversal properties of cross correlations in
  financial time series}.
\newblock Phys Rev Lett. 1999;83:1471--1474.

\bibitem{Drozdz-Grummer-Gorski-Ruf-Speth-2000-PA}
Dro\.~zd\. z S, Gr\"{u}mmer F, G\'{o}rski AZ, Ruf F, Speth J.
\newblock {Dynamics of competition between collectivity and noise in the stock
  market}.
\newblock Physica A. 2000;287(3-4):440--449.

\bibitem{Plerou-Gopikrishnan-Rosenow-Amaral-Guhr-Stanley-2002-PRE}
Plerou V, Gopikrishnan P, Rosenow B, Amaral LAN, Guhr T, Stanley HE.
\newblock {Random matrix approach to cross correlations in financial data}.
\newblock Phys Rev E. 2002;65:066126.

\bibitem{Tumminello-Lillo-Mantegna-2010-JEBO}
Tumminello M, Lillo F, Mantegna RN.
\newblock {Correlation, hierarchies, and networks in financial markets}.
\newblock J Econ Behav Org. 2010;75:40--58.

\bibitem{Song-Tumminello-Zhou-Mantegna-2011-PRE}
Song DM, Tumminello M, Zhou WX, Mantegna RN.
\newblock {Evolution of worldwide stock markets, correlation structure, and
  correlation based graphs}.
\newblock Phys Rev E. 2011;84:026108.

\bibitem{Dai-Xie-Jiang-Jiang-Zhou-2015-EmpE}
Dai YH, Xie WJ, Jiang ZQ, Jiang GJ, Zhou WX.
\newblock {Correlation structure and principal components in global crude oil
  market}.
\newblock Emp Econ. 2015;p. submitted.

\bibitem{SalesPardo-Guimera-Moreira-Amaral-2007-PNAS}
{Sales-Pardo} M, Guimer{\`a} R, Moreira AA, Amaral LAN.
\newblock {Extracting the hierarchical organization of complex systems}.
\newblock Proc Natl Acad Sci USA. 2007;104:15224--15229.

\bibitem{Lancichinetti-Fortunato-2012-SR}
Lancichinetti A, Fortunato S.
\newblock {Consensus clustering in complex networks}.
\newblock Sci Rep. 2012;2:336.

\bibitem{Baba-Shibata-Sibuya-2004-ANZJS}
Baba K, Shibata R, Sibuya M.
\newblock {Partial correlation and conditional correlation as measures of
  conditional independence}.
\newblock Aust N Z J Stat. 2004;46:657--664.

\bibitem{Kenett-Shapira-BenJacob-2009-JPS}
Kenett DY, Shapira Y, Ben-Jacob E.
\newblock {RMT assessments of the market latent information embedded in the
  stocks' raw, normalized, and partial correlations}.
\newblock J Prob Stat. 2009;2009:249370.

\bibitem{Kenett-Tumminello-Madi-GurGershgoren-Mantegna-BenJacob-2010-PLoS1}
Kenett DY, Tumminello M, Madi A, Gur-Gershgoren G, Mantegna RN, Ben-Jacob E.
\newblock {Dominating clasp of the financial sector revealed by partial
  correlation analysis of the stock market}.
\newblock PLoS One. 2010;5:e15032.

\bibitem{Kenett-Huang-Vodenska-Havlin-Stanley-2015-QF}
Kenett DY, Huang XQ, Vodenska I, Havlin S, Stanley HE.
\newblock Partial correlation analysis: Applications for financial markets.
\newblock Quant Finance. 2015;15(4):569--578.

\bibitem{Noh-2000-PRE}
Noh JD.
\newblock {Model for correlations in stock markets}.
\newblock Phys Rev E. 2000;61:5981--5982.

\bibitem{Kim-Jeong-2005-PRE}
Kim DH, Jeong H.
\newblock {Systematic analysis of group identification in stock markets}.
\newblock Phys Rev E. 2005;72:046133.

\bibitem{Bai-Ng-2002-Em}
Bai JS, Ng S.
\newblock {Determining the number of factors in approximate factor models}.
\newblock Econometrica. 2002;70(1):191--221.

\bibitem{Phillips-Sul-2007-Em}
Phillips PCB, Sul D.
\newblock {Transition modeling and econometric convergence tests}.
\newblock Econometrica. 2007;75(6):1771--1855.

\bibitem{Hodrick-Prescott-1997-JMCB}
Hodrick R, Prescott E.
\newblock {Postwar U.S. business cycles: An empirical investigation}.
\newblock J Money, Credit, and Banking. 1997;29(1):1--16.

\bibitem{Holmes-Grimes-2008-US}
Holmes MJ, Grimes A.
\newblock {Is there long-run convergence among regional house price in the UK?}
\newblock Urban Stud. 2008;45:1531--1544.

\end{thebibliography}

\end{document}